\documentclass[letterpaper,english ,preprint, aps]{revtex4-1}
\usepackage{lmodern}

\usepackage[T1]{fontenc}
\usepackage[latin9]{inputenc}
\usepackage{verbatim}
\usepackage{amsmath}
\usepackage{amssymb}
\usepackage{esint}

\makeatletter

\usepackage{graphicx}
\setcitestyle{super}

\makeatother

\usepackage{babel}
\begin{document}

\title{Optimising Graphene Visibility in van der Waals Heterostructures\\
}


\author{Thanmay~S.~Menon$^{1,*}$, Simli~Mishra$^{2}$, Vidhu~Catherine~Antony$^{3}$, Kiranmayi~Dixit$^{4,\dagger}$, Saloni~Kakkar$^{5}$, Tanweer~Ahmed$^{5}$, Saurav~Islam$^{5}$, Aditya~Jayaraman$^{5}$, Kimberly~Hsieh$^{5,*}$, Paritosh~Karnatak$^{5,\ddagger,*}$ and Arindam Ghosh$^{5,6}$}
\vspace{1.5cm}
\affiliation{$^1$ Undergraduate Programme, Indian Institute of Science, Bangalore 560012, India}
\affiliation{$^2$ Department of Physical Sciences, Indian Institute of Science Education and Research (IISER) Kolkata, Mohanpur 741246, Nadia, West Bengal, India}
\affiliation{$^3$ Department of Physics, Bangalore University, Bangalore 560 056, India}
\affiliation{$^4$ Department of Physics, Indian Institute of Technology Guwahati, Guwahati 781 039, Assam, India}
\affiliation{$^5$ Department of Physics, Indian Institute of Science, Bangalore 560012, India}
\affiliation{$^6$ Centre for Nano Science and Engineering, Indian Institute of Science, Bangalore 560012, India}

\thanks{These authors contributed equally to this work. Correspondence should be addressed to K.H. (kimberly@iisc.ac.in) or P.K. (paritosh@iisc.ac.in).}

\altaffiliation{\\$^\dagger$ Current address: Department of Physics and Astronomy, Purdue University, West Lafayette, IN 47907, USA}
\altaffiliation{\\$^\ddagger$ Current address: Department of Physics, University of Basel, Klingelbergstrasse 82, CH-4056 Basel, Switzerland}

\begin{abstract}
Graphene constitutes one of the key elements in many functional van
der Waals heterostructures. However, it has negligible optical visibility
due to its monolayer nature. Here we study the visibility of graphene
in various van der Waals heterostructures and include the effects
of the source spectrum, oblique incidence and the spectral sensitivity
of the detector to obtain a realistic model. A visibility experiment
is performed at different wavelengths, resulting in a very good agreement
with our calculations. This allows us to reliably predict the conditions
for better visibility of graphene in van der Waals heterostructures.
The framework and the codes provided in this work can be extended
to study the visibility of any 2D material within an arbitrary van
der Waals heterostructure.
\end{abstract}
\maketitle

\section*{Introduction}

The family of van der Waals materials has now expanded beyond graphene
and offers a wide range of material functionalities, such as semiconductors\citealp{Paul2016,Duan2015},
insulators\citealp{Bhattacharjee2017,Britnell2012,Duong2017}, superconductors\citealp{Kusmartseva2009,Xi2015}
and ferromagnets\citealp{Huang2017,Gong2017}. In addition, van der
Waals heterostructures are fabricated by stacking individual 2D materials
to obtain compound materials with novel functionalities\citealp{Liu2016,Georgiou2013,Roy2013_1}.
However, graphene is often difficult to locate through an optical
microscope during or after the heterostructure assembly due to its
near-optical transparency. While the visibility of graphene on SiO\textsubscript{2}/Si
substrate has been studied before\citealp{Roddaro2007,Blake2007,Casiraghi2007,Ni2007,Li2013rapid,Bing2018optical,Huang2019optical},
its visibility in various configurations of van der Waals heterostructures
has not yet been explored. In this paper, we develop a stepwise, robust
formalism to study the visibility of graphene in various heterostructures
and include practical considerations such as the details of the source
spectrum, oblique incidence which is necessary at high magnifications\citealp{Jessen2018,Bruna2009,Casiraghi2007,Bing2018optical,Katzen2018,Velicky2018,Bing2018optical,Huang2019optical},
and the spectral sensitivity of the detector. We perform the visibility
calculations for graphene-BN , graphene-BN-MoS\textsubscript{2},
graphene-MoS\textsubscript{2}-BN and graphene-MoS\textsubscript{2}
heterostructures on SiO\textsubscript{2} substrate and suggest conditions
for better visibility. In order to corroborate our theoretical calculations,
we experimentally determine the optical visibility of a graphene-BN-MoS\textsubscript{2}
and graphene-MoS\textsubscript{2} heterostructures at different wavelengths.
Our methods, codes and table of information can be employed to study
the visibility of most 2D van der Waals materials in an arbitrary
heterostructure configuration.

\section*{Methods}

For the theoretical calculations, a thin film interference model was
assumed where the reflection coefficients were calculated using Fresnel's
equations. Reflection coefficients $r_{i,s}$ and $r_{i,p}$ (for \textit{s} and \textit{p} polarisations respectively) for the interface between $i^{th}$ and $i+1^{th}$ layer (see Fig.
\ref{fig:1}b) can be written as\textsuperscript{\citealp{hecht2016optics,Born1999}}:
\begin{equation}
r_{i,s}=\frac{\tilde{n}_{i}cos(\theta_{i})-\tilde{n}_{i+1}cos(\theta_{i+1})}{\tilde{n}_{i}cos(\theta_{i})+\tilde{n}_{i+1}cos(\theta_{i+1})}\label{eq:1}
\end{equation}

\begin{equation}
r_{i,p}=\frac{\tilde{n}_{i+1}cos(\theta_{i})-\tilde{n}_{i}cos(\theta_{i+1})}{\tilde{n}_{i+1}cos(\theta_{i})+\tilde{n}_{i}cos(\theta_{i+1})}\label{eq:2}
\end{equation}

where $\tilde{n}_i$ $(\tilde{n}_{i+1})$ and $\theta_i$$(\theta_{i+1})$
corresponds to the refractive index and angle from the normal respectively
in the $i^{th}$ $(i+1^{th})$ layer. $\theta_i$ can be obtained
using Snell's law applied to the $i^{th}$ layer and is complex for
an absorbent material.

Often one or more layers in the stack are optically anisotropic, i.e.,
the in-plane and out-of-plane polarisation refractive indices are
not equal. For uniaxial crystals, the calculation of $r_{i,s}$ remains
trivial. However, the refractive index in the case of $p$-polarisation
is obtained by solving the following self-consistent equation\citealp{Born1999}
in $\theta$: %

\begin{equation}
\frac{1}{\tilde{n}_{p}^{2}}=\frac{\cos^{2}\theta}{\tilde{n}_{\parallel}^{2}}+\frac{\sin^{2}\theta}{\tilde{n}_{\perp}^{2}}\label{eq:3}
\end{equation}

where $n_{\parallel}$ and $n_{\perp}$ are the in-plane and out-of-plane
polarisation refractive indices respectively.

After calculating $r_i$ (for $s$- and $p$ polarisations), the reflectivity
can be obtained from the N-layer reflection formula, $\tilde{R}_{N}$
which is obtained recursively as follows\citealp{Anders65,knittl1976optics}:

If $\phi_{i}=4\pi\tilde{n}_{i}\cos\theta_{i}\,d_{i}/\lambda$ is the
phase shift due to the optical path difference in the $i^{th}$ layer
of thickness $d_i$,

\begin{equation}
\tilde{R}_{N}=R_{N}e^{i\delta_{N}}=\frac{r_{0}+\tilde{R}_{N-1}e^{-i\phi_{0}}}{1+r_{0}\tilde{R}_{N-1}e^{-i\phi_{0}}}\label{eq:4}
\end{equation}

where $\delta_{N}$ is the total phase acquired after transmission through $N$ layers, $r_0$ is the reflection coefficient of the topmost interface (see Fig. \ref{fig:1}b) and is given by \eqref{eq:1} or \eqref{eq:2}
depending on the polarisation of incident light, and $\tilde{R}_{N-1}$
is the reflection from $N-1$ layers which is computed by applying \eqref{eq:4}
to the substrate and using $r_{1}$ and $\phi_{1}$ instead of $r_{0}$
and $\phi_{0}$. This method is repeated for subsequent layers till
we reach $\tilde{R}_{0}=r_{N}$ which is the reflection from the interface
between the Nth layer and semi-infinite medium of silicon. The reflected
intensity from the entire stack is given by $I=\left|\tilde{R}_{N}\right|^{2}$.

Accounting for oblique incidence is especially important when viewing
the samples at large magnifications (especially 100$\times$) because of the
high numerical aperture ($\alpha_{NA}$) of the objective . To take this into
account, we assume that the incident beam has a gaussian profile\citealp{Bruna2009}
and integrate the reflected intensity for both polarisations over
angles from 0 to $\theta_M=a\sin(\alpha_{NA})$

\begin{equation}
I(\lambda)=\intop_{0}^{\theta_{m}}I(\theta,\lambda)\,e^{-\frac{2\rho^{2}}{\rho_{m}^{2}}}\tan\theta\,d\theta
\end{equation}

where $\rho=\tan\theta$ and $\rho_{m}=\tan\theta_{m}$ (see Fig.
\ref{fig:1}b)

In the visibility calculation, the normalisation constants for the
gaussian distribution cancel out eventually. As the incident beam
is unpolarised, the reflected intensity is the average of the reflected
intensities due to \textit{s} and \textit{p} polarisations.

The visibility (also known as Michelson's contrast)\citealp{InhwaJung2007,Michelson1927}
is defined as:

\begin{equation}
Visibility\,(in\:\%)=100\times\frac{I_{s}-I_{g}}{I_{s}+I_{g}}\label{eq:visibility}
\end{equation}

$I_{s}$ corresponds to the reflection from the substrate and $I_{g}$
refers to the reflection from the entire heterostucture. Positive
or negative value of visibility corresponds to graphene appearing
darker or lighter than the substrate, respectively.

Often the spectrum of the source and the spectral sensitivity of the
detector must also be included to obtain a more accurate value of
the visibility. For a typical RGB digital camera, if the red, green
and blue channel spectral sensitivities are $\omega_{R}(\lambda)$, $\omega_{G}(\lambda)$
and $\omega_{B}(\lambda)$ and the spectrum of the source is $S(\lambda)$,
then the reflected intensities picked up by the red, green and blue
channels are\citealp{Stigell2007,Jessen2018}:

\begin{equation}
\label{eq:source}
I_{j}=\intop_{0}^{\infty}I(\lambda)\,\omega_{j}(\lambda)\,S(\lambda)\:d\lambda
\end{equation}
where j=R, G, B.

By substituting these intensities in the visibility formula (\ref{eq:visibility}),
one can calculate the visibility for each colour channel. One should
also include the spectral dependence of lenses, mirrors, etc. for
more sensitive applications.

The dependence of graphene's visibility on wavelength for different
angles of incidence is shown in Figs. \ref{fig:1}c and \ref{fig:1}d
for graphene-SiO\textsubscript{2} and graphene-BN-SiO\textsubscript{2}
heterostructures respectively. Although there is a shift in the peak
position for both heterostructures\textbf{}, the relative shifts
in the height of the peaks are different for both heterostructure
configurations. Specifically, comparing between incidence at 60$^{\circ}$
and normal incidence, there is a greater difference in the height
of the peaks for graphene-BN-SiO\textsubscript{2} than for graphene-SiO\textsubscript{2}.
This explains why accounting for oblique incidence is important for
complicated heterostructure configurations. The codes employed in this work are available on GitHub~\citealp{GitHub}.

The heterostructures presented in this work were prepared by mechanically exfoliating the individual flakes of graphene,
BN and MoS\textsubscript{2} on separate SiO$_2$ substrates followed by micro-mechanical transfer technique using a polydimethylsiloxane (PDMS) stamp spin-coated with a transparent sacrificial polymer layer.\citealp{Zomer2011,Roy2013,Karnatak2016} Similarly prepared graphene-based heterostructures have been previously shown to exhibit relatively high mobilities when encapsulated by BN\citealp{Karnatak2016,Aamir2018} and have also found widespread functionality as ultra-sensitive photodetectors due to their efficient interlayer charge transfer\citealp{Roy2013,Roy2018}.

We have analysed and performed visibility calculations for three heterostructure
configurations: graphene-BN-SiO\textsubscript{2} and
graphene-MoS\textsubscript{2}-SiO\textsubscript{2}, and graphene-BN-MoS\textsubscript{2}-SiO\textsubscript{2}. The images
were taken using an Olympus UC30 camera mounted on an Olympus BX51
microscope through a 100$\times$ objective. The heterostructures were illuminated
using standard light emitting diodes (LEDs) of different wavelengths at a constant power of $2$~mW and the exposure time (1-2~sec) was set so as not to saturate any of the channels but also yield sufficient
signal-to-noise ratio.The images were split into three RGB channels and
were analysed using ImageJ software. The relative intensities of these
channels depend on the particular LED being used. The channel with
the maximum signal-to-noise ratio for each LED was used in the calculations. To minimise
errors due to uneven illumination, $I_{s}$ was recorded at a point
on the substrate close to the point on graphene where $I_{g}$ was
recorded and then visibility was calculated using \eqref{eq:visibility}.
This was done multiple times over the whole sample for the
same LED and the average visibility is plotted. The spectrum of the LEDs and the spectral sensitivity
of the camera (see supplementary information Fig. S1) were incorporated in the calculations
using \eqref{eq:source}. The experiment was performed using standard LEDs rather than the microscope's incandescent light source since it provides a better control over the choice and range of the source wavelength and also avoids complications that arise from digital image processing of a white-light image (such as white-balance), unique to each camera and its imaging software. The thickness of BN used in the stack was determined to be 11~nm using atomic force microscopy (AFM) and Raman spectroscopy was
used to verify that MoS\textsubscript{2} and graphene were monolayers.
We used a standard SiO\textsubscript{2}-Si substrate with an SiO\textsubscript{2}
thickness of 285~nm. We have used both the in-plane and out-of-plane
refractive indices of graphene\citealp{Djurisic1999} and MoS\textsubscript{2}\citealp{Hieu2018}
 and the refractive indices of BN\citealp{Gorbachev2011} and SiO\textsubscript{2}\citealp{Rodriguez-deMarcos2016}
(both of which have zero extinction coefficients) as well as that
of Si\citealp{PhysRevB.27.985}.

\section*{Results and discussion}

Figs. \ref{fig:2}(a) and \ref{fig:2}(c) show the experimental results
along with the theoretical calculations for two-layer graphene-BN
and graphene-MoS\textsubscript{2} heterostructures with SiO\textsubscript{2}
as substrate. The respective thicknesses were fixed at the experimentally determined values. To the right of Figs. \ref{fig:2}(a) and \ref{fig:2}(c) are the microscope images of the respective heterostructures illuminated with LEDs of different wavelengths. Although we have performed the visibility experiment using seven LEDs with wavelengths spread over the visible regime, we have shown only selected images in those colour channels with maximum signal-to-noise ratio. $\alpha_{NA}$ obtained from the best fit of the visibilities was $\sim0.88$ which is very close to the value of 0.9 provided by Olympus. We can see very close agreement between theory and experiment,
which suggests that our model for computing visibilities is sufficiently
accurate. Minor deviations of the experimental values from the theoretical
values may be due to the incident beam not being strictly gaussian
in nature. In Figs. \ref{fig:2}b and \ref{fig:2}d, we have presented
the results of our visibility calculations for the graphene-BN and
graphene-MoS\textsubscript{2} heterostructures as a function of varying substrate and underlayer thicknesses. It must be mentioned here that the calculations in Figs. \ref{fig:2}(a) and \ref{fig:2}(c) incorporated the spectrum of the LEDs in order to corroborate the results of the visibility experiment while the calculations in Figs. \ref{fig:2}b and \ref{fig:2}d have been performed assuming a typical $\alpha_{NA}$ of 0.9 and integrating over the spectrum of an incandescent light source of blackbody temperature 3100~K, the most common light source used in optical microscopes employed for the searching of suitable flakes of van der Waal materials for device fabrication. Although we have assumed the spectral sensitivity of our camera (SONY ICX252AQ CCD image sensor) in our calculations, it does not differ widely across different cameras\citealp{Jiang2013}.
Such a model gives us three values of visibility (one for each colour
channel). In these figures, we have plotted the maximum of the absolute
values of the three visibilities arising from red, green and blue
channels. From our experiments, we have arrived at an absolute visibility
threshold value of 2.5\% above which the graphene would be visible.
In this way, the phase space of BN-SiO\textsubscript{2} thicknesses
in Fig. \ref{fig:2}b can be divided into `islands of visibility'
(bounded by dashed white lines in the figure), where graphene would
be visible if the BN and SiO\textsubscript{2} thicknesses lie within
such an island. The same has been done in Fig. \ref{fig:2}d for graphene-MoS\textsubscript{2}
heterostructure. In these figures, we see that using lower thicknesses
of SiO\textsubscript{2} substrate results in higher visibility for
graphene in both graphene-BN and graphene-MoS\textsubscript{2} heterostructures.
Fig. \ref{fig:2}b also indicates that in a graphene-BN heterostructure
on SiO\textsubscript{2} (with a typical thickness of 300 nm), the
visibility of graphene reduces with increasing BN thickness. For BN
thicknesses above $\approx$ 42 nm, graphene is visible only at low
SiO\textsubscript{2} thicknesses and in the supplementary information (Fig. S2), we have
shown that a 380 nm double layer of polymethyl methacrylate (PMMA) spin-coated
on the heterostructure improves the visibility significantly for thicker
BN on 285 nm SiO\textsubscript{2}.

Fig. \ref{fig:3}a shows the experimental results and theoretical
calculations for graphene-BN-MoS\textsubscript{2} heterostructure (the right panels show the corresponding optical images under the illumination of various LEDs). We see good agreement between the theory and experiment even in a
three-layer heterostructure. The insets in Figs. \ref{fig:2}a,
\ref{fig:2}c and \ref{fig:3}a show that a model that considers only
normal incidence is inadequate in explaining the experimental results
and therefore, one must resort to an oblique incidence model to make
reliable predictions. Figs. \ref{fig:3}b and \ref{fig:3}c show visibility
calculations for graphene-BN-MoS\textsubscript{2} and graphene-MoS\textsubscript{2}-BN
heterostructures as a function of SiO\textsubscript{2} and BN thicknesses and assuming MoS\textsubscript{2} to be a monolayer. We observe that this figure looks similar to the graphene-BN
graph. This is because of the similarity between the refractive indices
of MoS\textsubscript{2} and BN especially in the red and green wavelength
ranges and because we are only considering monolayer MoS\textsubscript{2}.
Similar to the graphene-BN heterostructure, lower SiO\textsubscript{2}
thickness is recommended for better visibility of graphene.

Fig. \ref{fig:4} shows the maximum absolute visibility for various
heterostructure configurations on 285 nm of SiO\textsubscript{2}
along with the peak wavelengths. It also suggests various thicknesses
of van der Waals materials that can be used for viewing graphene on
285~nm SiO\textsubscript{2}. These values tend to lie in the green
wavelengths which our eyes are most sensitive to. This is not
a coincidence but a result of choosing SiO\textsubscript{2} thickness
to be 285~nm. From our calculations, we can conclude that effects of van der Waals materials under graphene are fundamentally no different when compared to a regular dielectric. Although the effect of the substrate dominates over non-metallic van der Waals materials, using an SiO$_2$ substrate of an arbitrary thickness to improve the visibility is often neither desirable nor practical. Instead, our aim in this paper is to provide a general framework and codes to allow the readers to calculate the visibility of any crystal within a heterostructure. In case of poor visibility, the readers may consider choosing an underlayer (say, BN) of appropriate thickness, if the calculations predict that this improves the visibility. From the figure, we also note that graphene-Bi\textsubscript{2}Se\textsubscript{3}
has very poor visibility. This is attributed to the high extinction coefficient
of Bi\textsubscript{2}Se\textsubscript{3}\citealp{Eddrief_2016}, due to the existence of its metallic surface states. Hence, we expect graphene's visibility to be adversely affected by metallic single- to few-layers. A similar reduction in contrast was already observed for graphene on gold substrates~\cite{Katzen2018,Velicky2018}.

In the supplementary information (Fig. S2), we have computed the visibility for other common
heterostructure configurations such as BN-graphene-BN, PMMA-graphene-BN,
PMMA-graphene-MoS\textsubscript{2} and a recently proposed photocatalyst, graphene-ZrS$_2$~\cite{Zhang2016} using the same method and have identified optimal thicknesses of different layers for maximum visibility.

In conclusion, we have studied the conditions for the optimal visibility of graphene
in van der Waals heterostructures. We have also performed experiments
to demonstrate the accuracy of our predictions. Our methods and codes
may be directly employed to calculate the visibility of a van der
Waals material in a heterostructure.

\subsection*{Acknowledgements}

The authors thank Am. Ghosh for use of a spectrometer and SERC, IISc for their computational facilities. The authors also acknowledge G. Singh and S. Majumder for useful discussions.

\newpage

\begin{center}
\vspace{20pt}
{\LARGE \bfseries Supplementary Information\\}
\vspace{50pt}
\end{center}

\maketitle

\renewcommand{\thefigure}{S\arabic{figure}}

\section{Table of references for refractive indices of various 2D materials}
\begin{center}
\begin{tabular}{|c|c|c|}
\hline 
S.I No. & Material & Reference\tabularnewline
\hline 
\hline 
1 & Graphene & Djuri¨ic et al., \textit{J. Appl. Phys.} 85, 7404 (1999)\tabularnewline
\hline 
2 & BN & Gorbachev et al., \textit{Small} 7, 465 (2011)\tabularnewline
\hline 
3 & MoS\textsubscript{2} & Hieu et al., \textit{Superlatt. Microstruct.} 115, 10 (2018)\tabularnewline
\hline 
4 & Bi\textsubscript{2}Se\textsubscript{3} & Eddrief et al., \textit{J Phys. D Appl. Phys.} 49, 505304 (2016)\tabularnewline
\hline 
5 & SiO\textsubscript{2} & Rodríguez-de Marcos et al., \textit{Opt. Mater. Express} 6, 3622
(2016)\tabularnewline
\hline 
6 & Si & Aspnes et al., \textit{Phys. Rev. B} 27, 985 (1983)\tabularnewline
\hline 
\end{tabular}
\par\end{center}

\section{Experimental Details}

The visibility of graphene-BN and graphene-BN-MoS\textsubscript{2}
heterostructure stacks were measured on different parts of the same sample. Figs. S1(a)
and (b) show the images of graphene-BN-MoS\textsubscript{2} and
graphene-MoS\textsubscript{2} sample under different light sources.
The light sources used were LEDs of different colours and their intensities
were calibrated to ensure uniform intensity of the LEDs. We use a gaussian beam approximation for the LEDs since the incident light beam is narrower than the input aperture of the objective lens (some high-end microscope light sources provide uniform illumination). The spectra of the LEDs, recorded using an Ocean optics spectrometer,
is given in Fig. S1(e). The images were taken using an Olympus UC30
camera, which was equipped with a SONY ICX 252 AQ CCD image sensor.
This sensor's spectral sensitivity (obtained from the sensor's datasheet)
is given in Fig. S1(f). These images were then split into RGB channels
and the channel with maximum signal-to-noise ratio was analysed using
ImageJ software. Figs. S1(c) and (d) show the same sample under an ordinary
white light source (blackbody of temperature $3100$~K).

\raggedright

\section{Results for other heterostructures}

Figs. S2(a) and S2(b) show the effect of polymethyl methacrylate
(PMMA) polymer on the maximum absolute visibility of graphene-BN and graphene-MoS\textsubscript{2} heterostructures.
A standard double layer of PMMA-495K (A4) followed by  PMMA-950K (A4), spin-coated at $\sim4500$~rpm, gives a thickness of $\sim380$~nm~\cite{2001Microchem}. We have assumed the same refractive indices for both
coatings of PMMA~\cite{2009AcPPA.116..585S}. We see that in case of graphene-BN heterostructure, PMMA increases the range of BN thicknesses which are optically visible under a microscope,
especially at $\sim300$~nm and $\sim90$~nm thicknesses of SiO\textsubscript{2}. However,
in the case of graphene-MoS\textsubscript{2} heterostructure, at
285 nm SiO\textsubscript{2} thickness, graphene is just barely visible
on monolayer MoS\textsubscript{2}.

Fig. S2(c) shows maximum absolute visibility of BN-graphene-BN heterostructure
where the thickness of the top layer of BN is fixed at 10~nm and the
thickness of the bottom BN layer below graphene is allowed to vary. This
figure shows features very similar to that of graphene-BN heterostructure
except that the maximum BN thickness for graphene to be visible is lesser
here. Fig S2 (d) shows the maximum absolute visibility map of graphene-Bi\textsubscript{2}Se\textsubscript{3}
heterostructure. Here the maximum absolute visibility is calculated
from 10~nm-thick Bi\textsubscript{2}Se\textsubscript{3} onwards
as there is evidence that the refractive index of Bi\textsubscript{2}Se\textsubscript{3}
abruptly changes below 10~nm~\cite{Vajner_2016}. We can see that
in the range of thicknesses we have considered for Bi\textsubscript{2}Se\textsubscript{3}, the visibility of graphene is very poor. This is attributed to the high extinction coefficient of Bi\textsubscript{2}Se\textsubscript{3}, due to its metallic surface states.

From Fig. S3a, we can see that the maximum visibility of graphene on monolayer ZrS$_2$ (SiO$_2$ thickness being 285~nm) is close to 6$\%$ at wavelength $\lambda \approx 527$~nm. This is on par with graphene's visibility on BN which would be very helpful during device fabrication. The refractive index and monolayer thickness of ZrS$_2$ were taken from Ref.~\cite{Vu2019} and Ref.~\cite{Wang2016} respectively.

\newpage
\begin{figure*}

\includegraphics{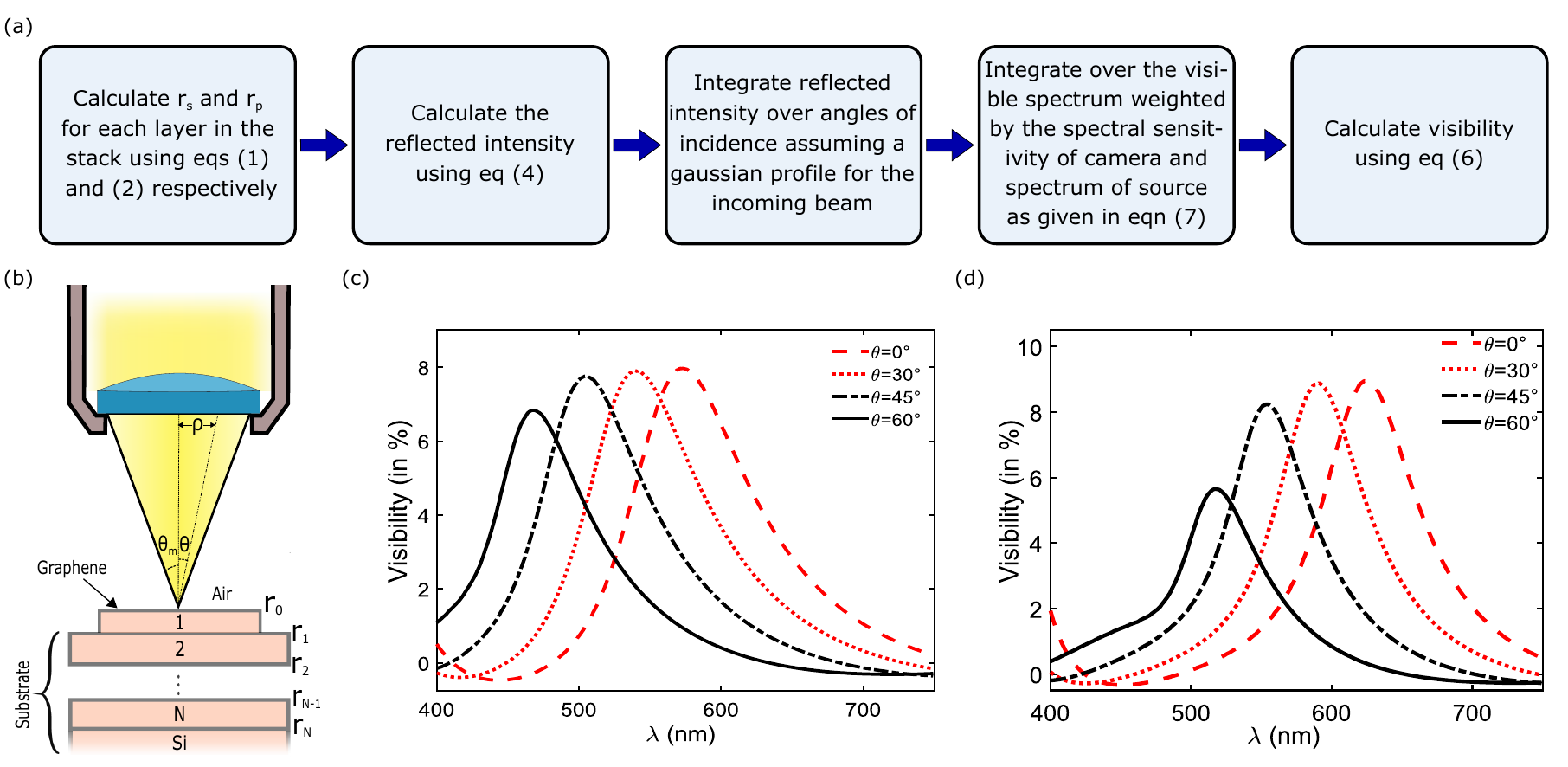}
\caption{\label{fig:1}\textbf{(a)} Block diagram showing how visibility is calculated for an arbitrary van der Waals heterostructure \textbf{(b)} A schematic of the $N$-layer stack illuminated by light passing through the objective. \textbf{(c,d)} Plot of visibility vs wavelength of \textbf{(c)} graphene on SiO\textsubscript{2}-Si substrate and \textbf{(d)} graphene on BN-SiO\textsubscript{2}-Si substrate for different angles of incidence.}

\end{figure*}

\begin{figure*}
\includegraphics{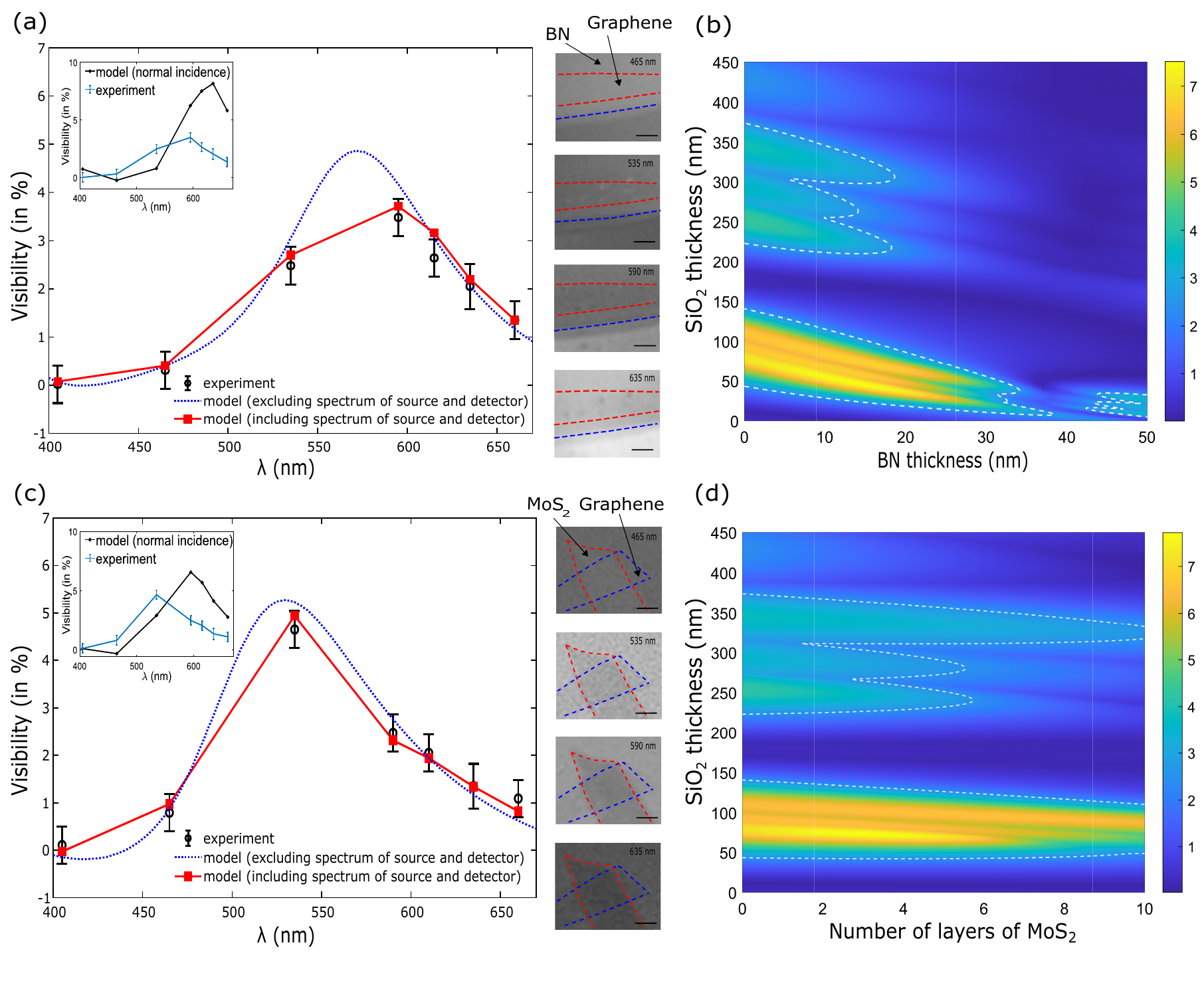}
\caption{\label{fig:2} \textbf{(a,c)} Visibility as a function of wavelength showing experimental and theoretical plots for \textbf{(a)} graphene-BN  \textbf{(c)} graphene-MoS\textsubscript{2} heterostructures. The panels to the right of \textbf{(a)} and \textbf{(c)} are the optical micrographs of the respective samples under the illumination of LEDs of different wavelengths (namely $465$~nm, $535$~nm, $590$~nm and $635$~nm) converted to greyscale. The scale bar at the bottom right is 2$\, \mu$m. Color plots of the maximum of absolute visibilities of the three channels as a function of \textbf{(b)} SiO\textsubscript{2} thickness and BN thickness \textbf{(d)} SiO\textsubscript{2} thickness and number of layers of MoS\textsubscript{2}. White dashed lines are contours drawn at 2.5\% absolute visibility.}
\end{figure*}

\begin{figure*}
\includegraphics{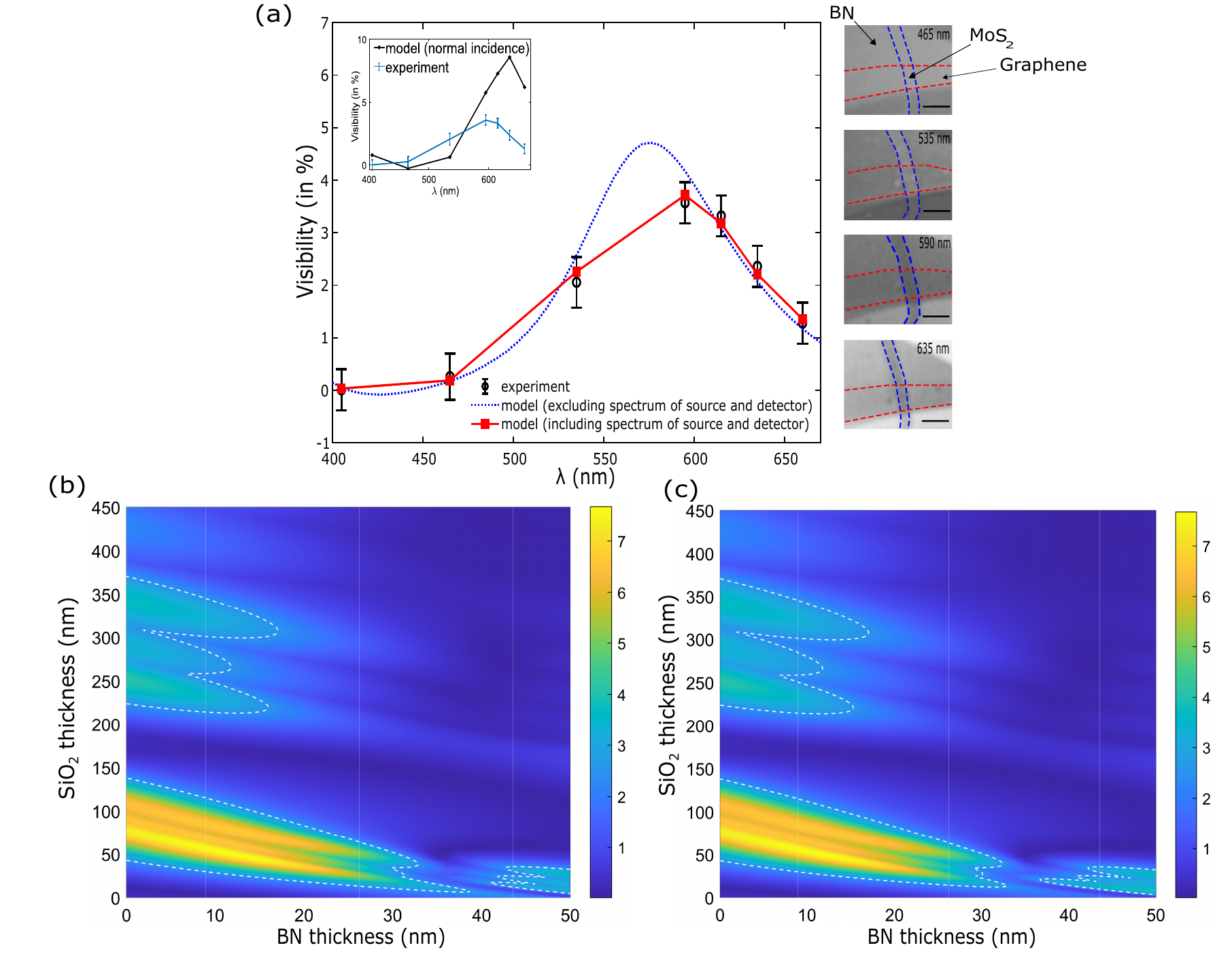}
\caption{\label{fig:3} \textbf{(a)} Visibility as a function of wavelength for graphene-BN-MoS\textsubscript{2}heterostructure showing experimental and theoretical plots. The panels to the right of \textbf{(a)} are optical micrographs under the illumination of LEDs of different wavelengths (namely $465$~nm, $535$~nm, $590$~nm and $635$~nm) converted to greyscale. The scale bar at the bottom right is 2$\, \mu$m. Color plots of the maximum of absolute visibilities of  \textbf{(b)} graphene-BN-MoS\textsubscript{2} and \textbf{(c)} graphene-MoS\textsubscript{2}-BN heterostructures. White dashed lines are contours drawn at 2.5\% absolute visibility.}
\end{figure*}

\begin{figure}
\includegraphics{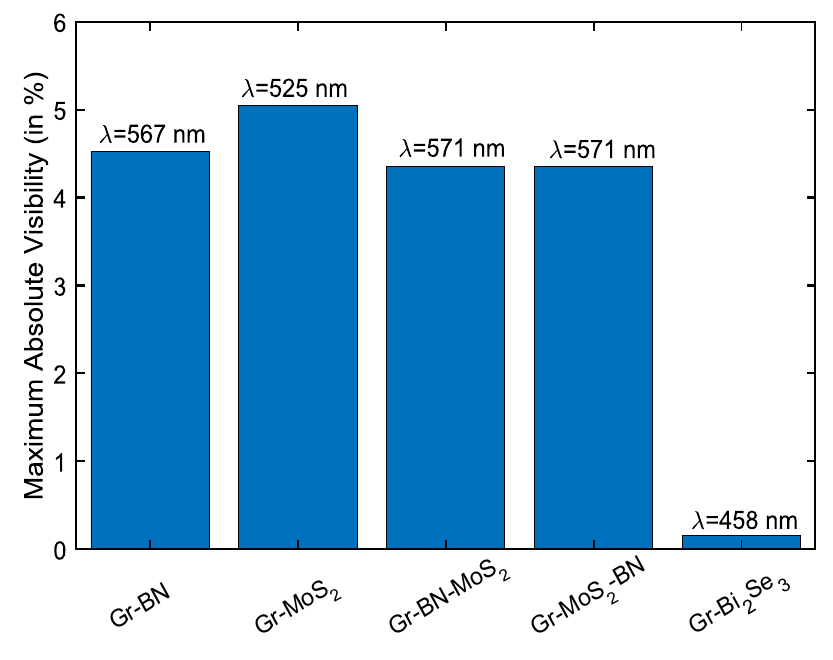}
\caption{\label{fig:4}Maximum absolute visibilities and their peak wavelengths for various heterostructures. BN thickness is assumed to be 11 nm and MoS\textsubscript{2} is assumed to be monolayer in all heterostructures. Bi\textsubscript{2}Se\textsubscript{3} thickness is assumed to be 10 nm.}
\end{figure}

\begin{figure*}
\includegraphics[width=1\linewidth]{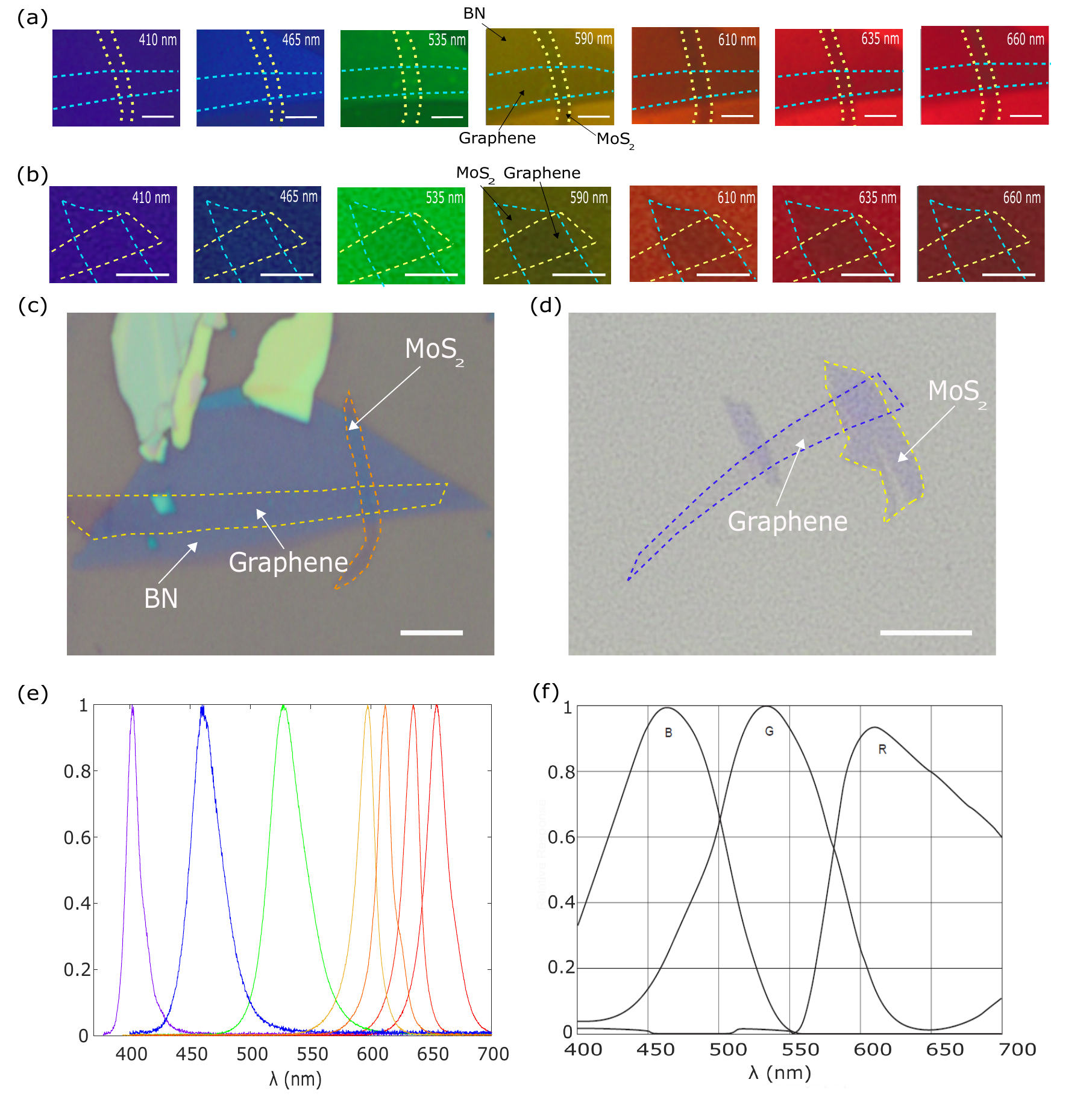}
\caption{Images of \textbf{(a)} graphene-BN-MoS2 and \textbf{(b)} graphene-MoS\textsubscript{2}
heterostructures illuminated by LEDs of different wavelengths. The scale bar at the
bottom right of these two sets of images are 2~$\mu$m in length. \textbf{(c)}
Graphene-BN-MoS\textsubscript{2} and \textbf{(d)} graphene-MoS\textsubscript{2}
heterostructures illuminated by a typical microscope light source.
The scale bar at the bottom right of these two images are 5~$\mu$m in
length. \textbf{(e)} Spectrum of the LEDs and \textbf{(f)} spectral
sensitivity of the camera used in the experiment.}
\end{figure*}

\begin{figure*}

\includegraphics[width=1\linewidth]{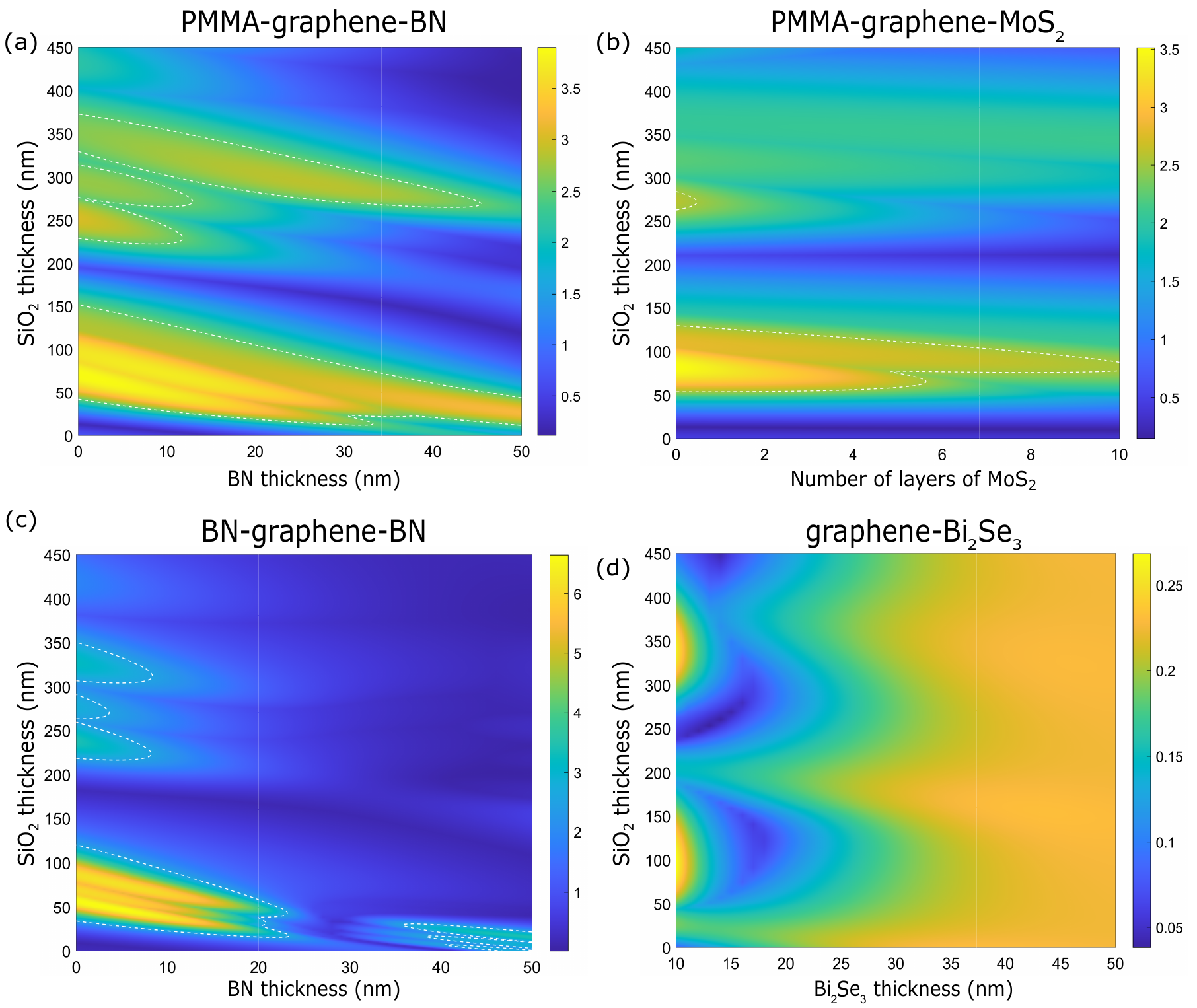}
\caption{Color plots of maximum absolute visibilities of the three channels
as a function of SiO\textsubscript{2} thickness and BN thickness
for (a) PMMA-graphene-BN and (c) BN-graphene-BN heterostructures. Color plots of maximum absolute visibilities of the three channels for (b) PMMA-graphene-MoS\textsubscript{2} heterostructure as a function of SiO\textsubscript{2} thickness and number of MoS\textsubscript{2} layers and (d) graphene-Bi\textsubscript{2}Se\textsubscript{3} heterostructure as a function of SiO\textsubscript{2} thickness and Bi\textsubscript{2}Se\textsubscript{3} thickness.
Thickness of PMMA in (a) and (b) is assumed to be 380 nm (double layer of PMMA spin-coated at 5000 RPM) and in BN-graphene-BN heterostructure, the thickness of the top BN is assumed to be 10~nm.}
\end{figure*}

\begin{figure*}
\includegraphics[width=1\linewidth]{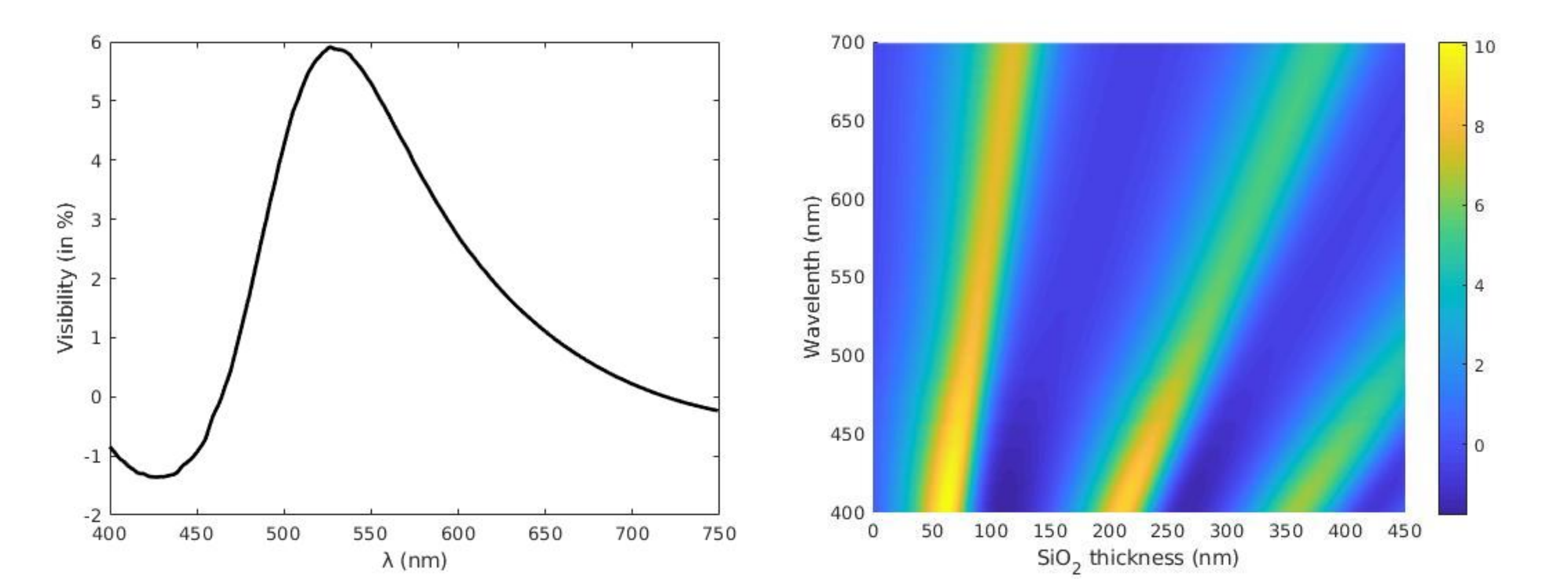}
\caption{(a) Visibility of graphene-ZrS$_2$ heterostructure as a function of wavelength of light for a fixed SiO$_2$ thickness of 285~nm. (b) Color plot of maximum absolute visibilities of the three channels for graphene-ZrS$_2$ heterostructure as a function of SiO\textsubscript{2} thickness and wavelength of light. Both graphene and ZrS$_2$ are assumed to be monolayers.}
\end{figure*}

\end{document}